\documentclass[aps,pra,twocolumn,amsmath,amssymb]{revtex4}%
\usepackage{graphicx}
\usepackage{epsfig}
\usepackage{subfigure}
\usepackage{graphics,color}
\usepackage[none]{hyphenat}
\usepackage{hyperref}
\usepackage{amsbsy}
\usepackage{mathtools}
\usepackage{amsmath}
\usepackage{bm}
\usepackage{caption}

\begin{document}
\title{High success standard quantum teleportation using entangled coherent state and two-level atoms in cavities}
\author{Ravi Kamal Pandey }
\email{ravikamalpandey@gmail.com}
\author{Ranjana Prakash}
\email {prakash_ranjana1974@rediffmail.com}
\author{Hari Prakash}
\email {prakash_hari123@rediffmail.com}
\affiliation{Physics Department, University of Allahabad, Prayagaj 211002, India.}
\date{\today}
\begin{abstract}
We propose here a new idea for quantum teleportation of superposed coherent state which is not only almost perfect, in principle, but also feasible experimentally. We use entangled resource  $\sim |\alpha,\frac{\alpha}{\sqrt{2}}\rangle-|-\alpha,-\frac{\alpha}{\sqrt{2}}\rangle$ in contrast with the usual $\sim |\alpha,\alpha\rangle-|-\alpha,-\alpha \rangle$ (both states unnormalized). Bob receives state which is then superposition of the states $|\pm \frac{\alpha}{\sqrt{2}}\rangle$ . Bob mixes these with even or odd coherent states involving superposition of states $|\pm \frac{\alpha}{\sqrt{2}}\rangle$ to obtain a two-mode state which is one of $\sim |I,0\rangle \pm |0,I\rangle$, $|I\rangle$ being the information state. Bob then obtains the teleported information by using interaction of one of these modes in two cavities with resonant two-level atoms. This scheme results in average fidelity of $\simeq 0.95$ for $|\alpha|^2 \simeq 10$, which increases with $|\alpha|^2$ and tends to 1 asymptotically, varying as $1-\frac{\pi^2}{16|\alpha|^2}+\frac{\pi^2(\pi^2+8)}{256|\alpha|^4}$  for large values of $|\alpha|^2$.

\end{abstract}

\maketitle
\section{Introduction}
\label{sec:1}
Quantum teleportation is faithful transmission of quantum information between two parties by sharing an appropriate entangled resource and a classical communication channel. Since its first theoretical proposal by Bennet et al. \cite{bennett1993teleporting} for teleporting the quantum state of a two-level system (spin state of spin-1/2 particles or polarization state of a photon), it has been realized experimentally in variety of physical systems including optical modes, photonic qubits, trapped ion etc.\cite{pirandola2015advances}. Quantum communication relies on the ability to perform efficient quantum teleportation over arbitrarily large distances. However, the pre-requisite demands to establish entanglement between the sender and receiver, which is limited by various noise factors. Entangled state passing through a quantum channel degrades naturally into a mixed one thereby reducing the fidelity of teleported state. To overcome losses, schemes such as quantum repeaters and relays can be employed for entanglement purification\cite{briegel1998quantum,sangouard2010quantum,
sangouard2011quantum}. Furthermore, performing Bell state measurement, deterministically, is important to achieve good success probability. In the context of two two-level bosonic states, it had been shown by L{\"u}tkenhaus et al. \cite{lutkenhaus1999bell} that a never-failing Bell state measurement is not possible, and in fact, using only linear optical devices such as phase shifters, beam splitters, and photo-detectors only two out of four Bell states can be discriminated.

Recently, quantum teleportation using hybrid quantum channel has been reported in which encoding of information can be changed from discrete variable to continuous variable or vice versa \cite{ulanov2017quantum,lee2013near}. Such a scheme has the advantage that whereas discrete systems, like trapped atoms, have longer information holding capacity, continuous variable shows greater robustness against environmental de-coherence and thus suitable for transmitting information over noisy channel \cite{hirota2001entangled,park2010entangled}. For all the progress that has been made in past three decades, the motivation to go further remains unaltered, finding physical systems which are most efficient and, at the same time, experimentally feasible.

Coherent states of radiation field,\cite{glauber1963coherent}, which are eigenstates of photon annihilation operator and are most close to classical noiseless field, form a valuable tool in performing various quantum information processing tasks. The coherent states are in general non-orthogonal, and in fact, for two coherent state $|\alpha\rangle$ and $|\beta\rangle$ (where $\alpha$ and $\beta$ are, in general, complex numbers) their overlap is given by,
\begin{equation}
|\langle\alpha|\beta\rangle|^{2}=\exp(-|\alpha-\beta|^{2}).
\label{eqn:1} 
\end{equation}
For $\beta=-\alpha$, $|\langle\alpha|-\alpha\rangle|^{2}=\exp(-2|\alpha|^{2})< 10^{-3}$ for $|\alpha|^{2}\geq 3$. Therefore, for moderately large coherent amplitudes, we can have a close correspondence between logical qubits $|0\rangle, |1\rangle$ and phase opposite coherent states $|\alpha\rangle, |-\alpha\rangle$, enabling us to encode and manipulate information in their superposition. Various schemes were proposed for the generation of such superposition states and also their entangled counterparts using non-linear interactions, photon substraction from squeezed vacuum states, mixing of squeeze vacuum with a coherent light etc., with appreciable large size \cite{yurke1986generating,sanders1992entangled,takahashi2008generation,
marek2008generating,huang2015optical,mikheev2019efficient} . Proposals for generating freely travelling multi-partite resources such as GHZ, W and cluster entangled coherent state (ECS) were also proposed \cite{an2009cluster,jeong2006greenberger}. Realizing the advantages that superposed coherent state (SCS) can offer, with regards to its robustness over noisy channel and the fastest mode of transmitting information, the field is since then, ever growing. \cite{jeong2001quantum,
jeong2002efficient,
ralph2003quantum,sanders2012review}. 

Jeong et al. studied quantum information processing using mixed-entangled coherent state, further extending it for efficient quantum computing. \cite{jeong2001quantum,jeong2002efficient}. Van Enk et al. \cite{van2001entangled} presented a scheme of achieving quantum teleportation of a single qubit information, encoded in phase opposite coherent state of the form, 
  \begin{equation}
  \label{eqn:2}
  |I\rangle=\epsilon_{+}|\alpha\rangle+\epsilon_{-}|-\alpha\rangle
  \end{equation} 
using ECS as quantum channel, similar to Standard Quantum Teleportation scheme of Bennet et al. \cite{bennett1993teleporting} for the teleportation of single qubit state living in 2-d Hilbert space. Although, the success probability of such a scheme was shown to be only $1/2$. This has been generalized for teleportation of ECS by Wang \cite{wang2001quantum}, with success probability of $1/2$. The cause of failure is the inability for the receiver to find a valid unitary transformation that may change $\epsilon_{+}|\alpha\rangle-\epsilon_{-}|-\alpha\rangle $ to $\epsilon_{+}|\alpha\rangle+\epsilon_{-}|-\alpha\rangle$ which correspond to making a $Z_c=|\alpha\rangle\langle\alpha|-|-\alpha\rangle\langle-\alpha|$ operation on SCS. $Z_c$ is a non-unitary operation due to non-orthogonality of coherent states, and it may become unitary in the limit when $|\alpha|^{2}\rightarrow\infty$. Jeong et al. \cite{jeong2002efficient} showed that an arbitrary rotation about z-axis can be obtained by displacement operator, however, such a displacement is rather hard to implement in practice requiring the state to be mixed with intense coherent beam ($|\alpha|^{2}\rightarrow\infty$) using highly reflecting beam splitter \cite{paris1996displacement}. Ralph et al. \cite{ralph2003quantum} discussed the implementation of such single flip or $Z_c$ gate operation using quantum teleportation and making $X$ gate error correction. Although, $X$ operation for SCS can be implemented by simply passing the mode through a $\pi$ phase shifter, this scheme fails half of the time. Cheong et al. \cite{cheong2004near} presented a teleportation scheme for SCS in which the required $Z_c$ operation is approximately obatined by interacting the mode with two-level atom in a cavity. N Ba An \cite{an2003teleportation} discussed a scheme of quantum teleportation of SCS within a network, also with success probability $1/2$. Prakash et al. \cite{prakash2007improving} presented a modified scheme for the teleportation of SCS, where it is shown that by adopting a photon counting strategy that can discriminate between a zero, non-zero even and an odd photon count, almost perfect teleportation for appreciable coherent state amplitude can be achieved. This scheme circumvents the problem of performing non-unitary $Z_c$ transformation by performing a unitary $Z$ operation on even ($|EVEN,\alpha\rangle$) and odd ($|ODD,\alpha\rangle$) coherent states, 
\begin{equation} 
\begin{gathered}
\label{eqn:4}    
        |EVEN,\alpha\rangle=[\sqrt{2(1+x^{2})}]^{-1}(|{\alpha}\rangle+|-{\alpha}\rangle)
       \\
 |ODD,\alpha\rangle=[\sqrt{2(1-x^{2})}]^{-1}(|{\alpha}\rangle-|-{\alpha}\rangle, 
 \end{gathered}  
  \end{equation}
which forms an orthogonal basis \cite{dodonov1974even}. This photon counting strategy has a great theoretical advantage over all the previously existing schemes that involve coherent states and it has been widely used for obtaining almost perfect success for many other quantum information processing task using ECS as a resource \cite{prakash2009entanglement,prakash2009increase,mishra2010teleportation,prakash2009swapping,
prakash2010almost,prakash2011quantum,Prakash2019,pandey2019controlled}. 
However, the experimental feasibility of the required $Z$ unitary operations demands a change between $|EVEN,\alpha\rangle\leftrightarrow|ODD,\alpha\rangle$ which seems difficult to achieve and so far no one has shown how this would be done experimentally. It is, therefore, important to devise a scheme which leads to teleportation of SCS with good success rate and fidelity, and at the same time permits its experimental feasibility using optimal resources. 
 
We in our work show that an almost perfect teleportation of SCS is obtainable, not only in principle, but which is also experimentally feasible. We shall show that by making or not making a phase-shift of $\pi$ in the state Bob receives and then adding extra step of mixing this received state with a preferred cat state (odd or even coherent state, conditioned to the classical information of sender) gives an almost perfect teleportation of SCS.
\section{Feasible scheme of quantum teleportation of SCS: getting two mode state $\sim |I,0\rangle \pm |0,I\rangle$ at Bob's stataion}
\label{sec:2}
\begin{figure}[t]
  \includegraphics[width=\linewidth]{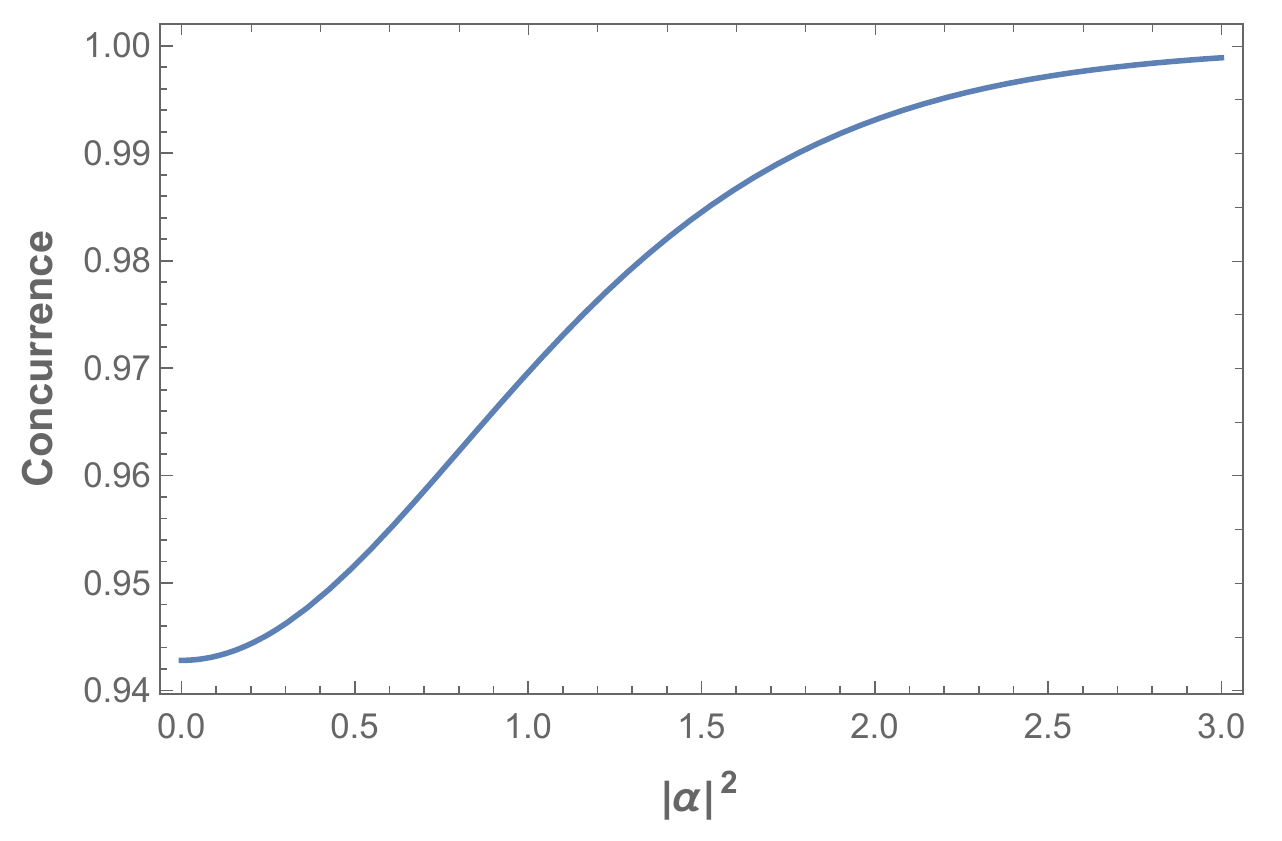}
  \caption{Variation of concurrence of the entangled channel, $|E\rangle_{1,2}$, used in our scheme with respect to mean photons in the coherent state, $|\alpha|^{2}$. The concurrence becomes almost unity for $|\alpha|^{2}\geqslant 2$.}
  \label{fig:1}
\end{figure}
\begin{figure}
  \includegraphics[width=\linewidth]{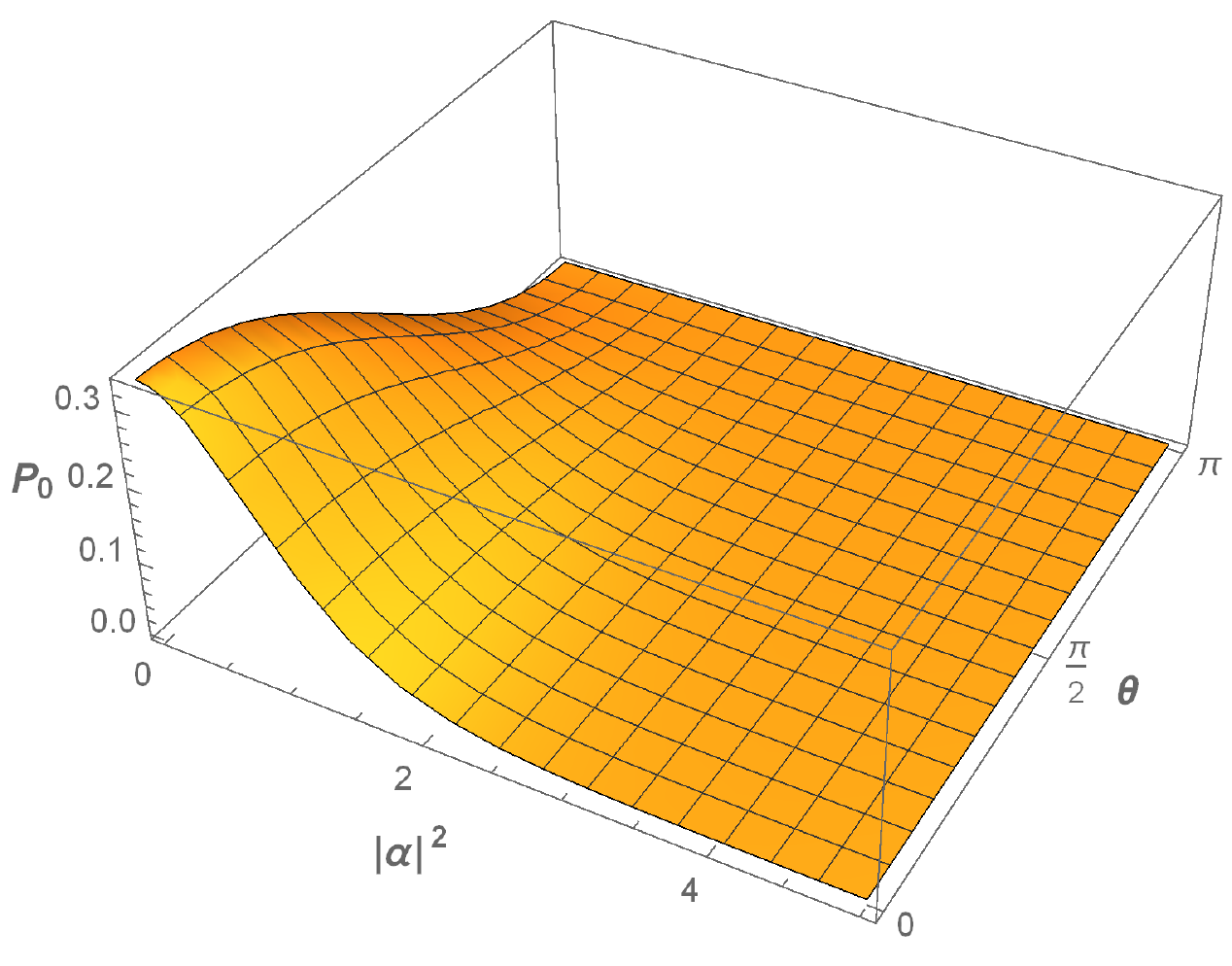}
  \caption{Variation of probability of occurrence for case (i)  with respect to information parameter $\theta$ and average photon number in the coherent state $|\alpha|^{2}$. The probability decreases rapidly to become vanishingly small for $|\alpha|^{2}\geq 3$.}
   \label{fig:2}
\end{figure} 
\begin{figure*}[!htbp]
\subfigure{
\label{fig:3a} 
\begin{minipage}[b]{0.45\linewidth}
\centering \includegraphics[width=3.32in]{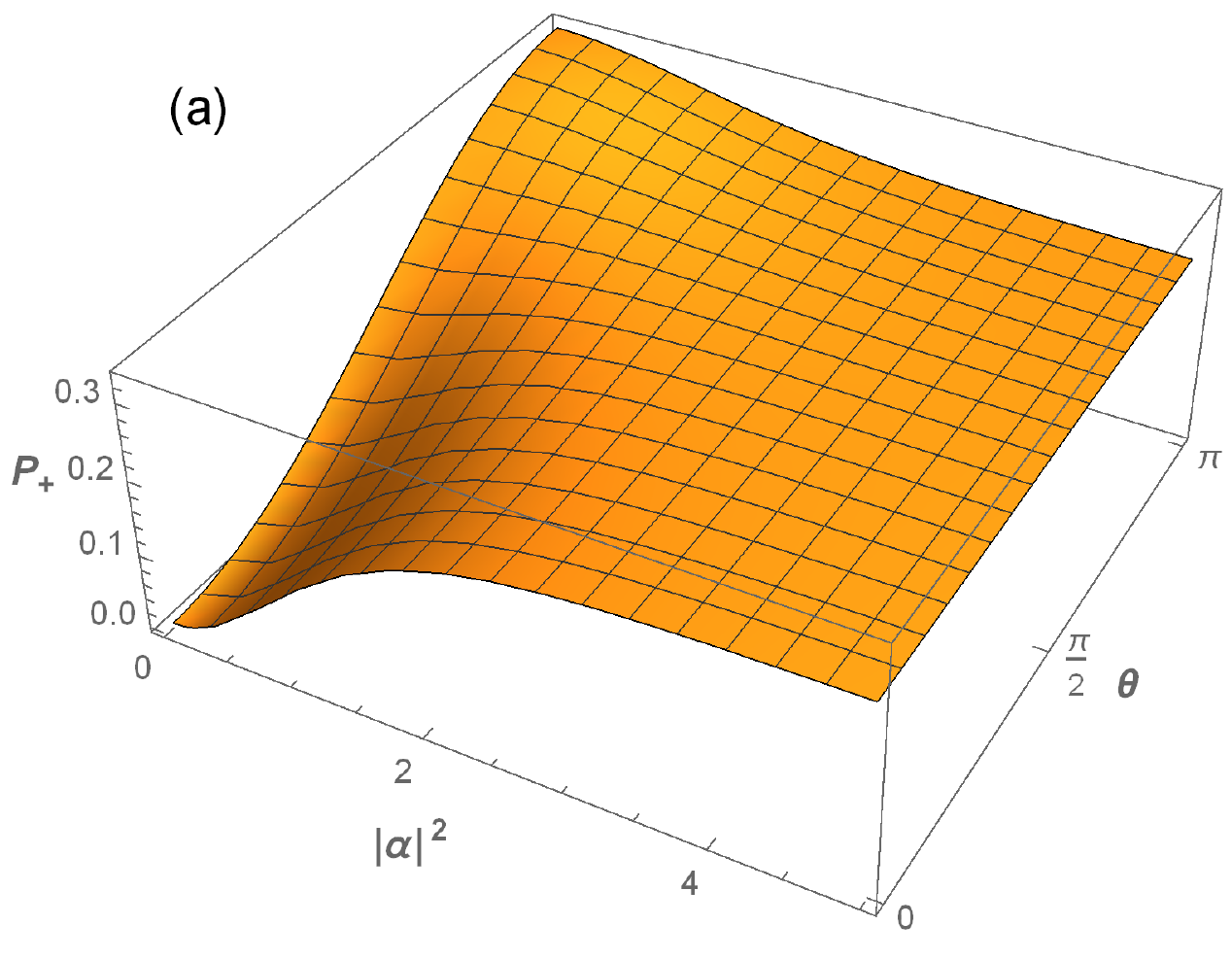}
\end{minipage}}%
\hspace{.12in}
\subfigure{
\label{fig:3b} 
\begin{minipage}[b]{0.45\linewidth}
\centering \includegraphics[width=3.32in]{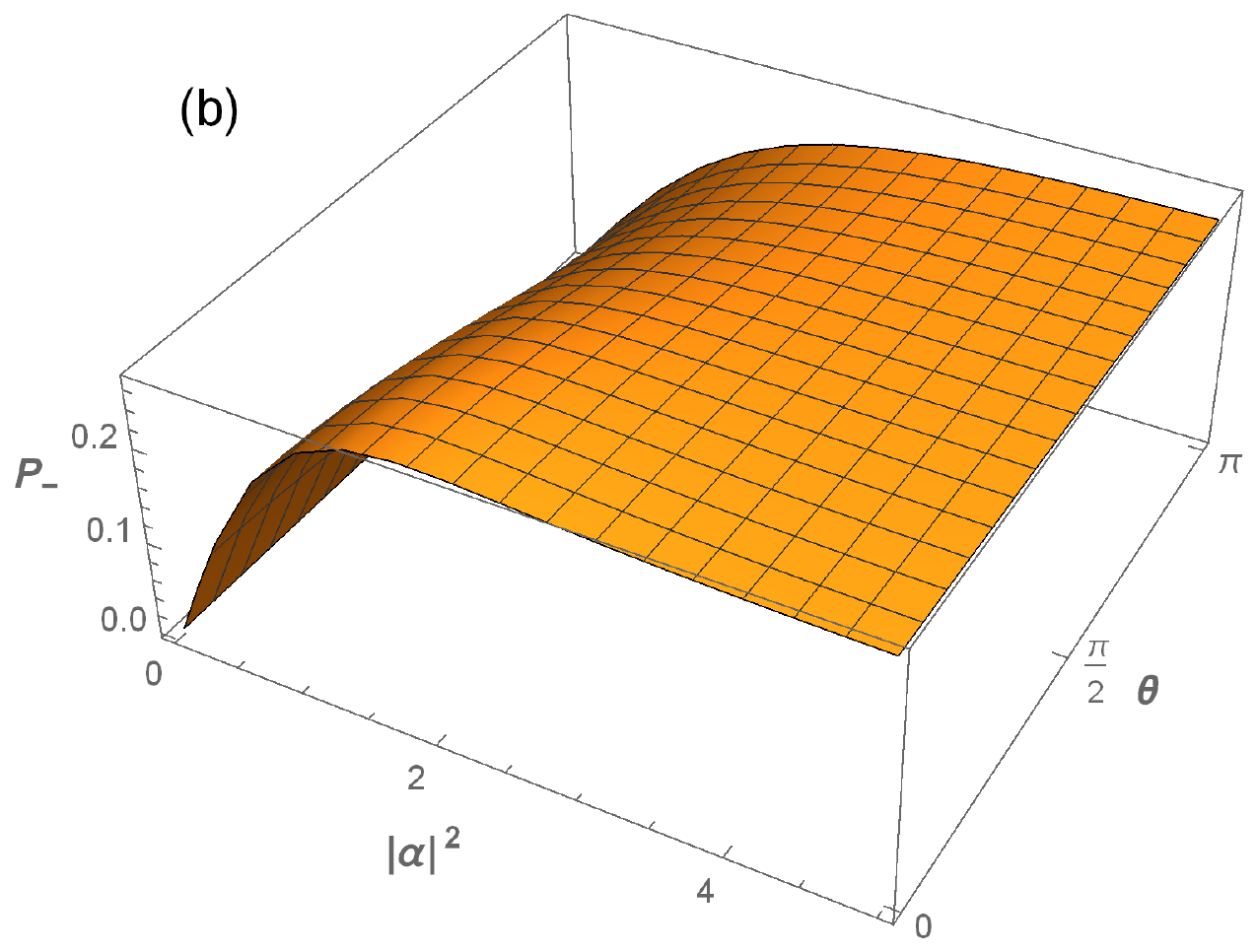}
\end{minipage}}
\caption{(a) and (b) shows the variation of probability of occurrence for cases (i) (same for case (ii)) and case (iii) (same for case (iv)) respectively, with respect to information parameter $\theta$ and average photon number in the coherent state $|\alpha|^{2}$. The probabilities converges to a constant value of 0.25 when there is appreciable mean photons in the coherent state.}
\label{fig:3} 
\end{figure*}
Let Alice possess a single qubit information state encoded in phase opposite coherent states given by Eq. \ref{eqn:2}, existing in mode 0. We can equivalently write the information state in orthogonal Cat state basis by using Eq. 3 in the form,
\begin{equation}
\label{eqn:5}
|I\rangle=A_{+}|EVEN,\alpha\rangle+A_{-}|ODD,\alpha\rangle,
\end{equation}
with $A_{\pm}=\sqrt{(1\pm x^{2})/2}(\epsilon_{+}+\epsilon_{-})$.  While preserving the normalization, $|A_{+}|^{2}+|A_{-}|^{2}=1$, one can write $A_{+}/A_{-}=e^{i\phi}\tan(\theta/2) $, using Bloch representation of a qubit. We can also expand information in terms of Fock basis $|n\rangle$ as, 
\begin{equation}
\label{eqn:6}
|I\rangle_{0}=\sum_{n=0}^{\infty}p_{In}|n\rangle_0
\end{equation}
where, $p_{In}=\sqrt{x}[ \epsilon_+ +(-1)^n\epsilon_-]\alpha^{n}/n!$ is the expansion coefficient. Normalization demands,
\begin{equation}
\label{eqn:7}
|\langle I|I\rangle|^2=\sum_{n=0}^{\infty} P_{In}=|\epsilon_{+}|^{2}+|\epsilon_{-}|^{2}+x(\epsilon_{+}^{*}\epsilon_{-}+\epsilon_{+} \epsilon_{-}^{*})=1
\end{equation} 
where, $P_{In}=|p_{In}|^{2}$ is the probability of counting $n$ photons in $|n\rangle$. Alice and Bob shares mode 1 and 2 of the entangled state,
\begin{equation}
\label{eqn:8}
|E\rangle_{1,2}=\frac{1}{\sqrt{2(1-x^{3}]}}\left(\left|\alpha,\frac{\alpha}{\sqrt{2}}\right\rangle_{1,2}-\left|-\alpha,-\frac{\alpha}{\sqrt{2}}\right\rangle_{1,2}\right)
\end{equation}
respectively. The reason for selecting such an entangled channel shall become clear in a while. It should be noticed that $|E\rangle_{1,2}$ is not a maximally entangled state in terms of its concurrence \cite{wootters1998entanglement}, due to difference in coherent amplitudes of the two modes and this is in contrast to the ECS used usually, which has equal amplitude \cite{van2001entangled,jeong2001quantum,prakash2007improving}and exactly 1-e bit entanglement for all $\alpha\neq 0$ \cite{hirota2001entangled}. However, its concurrence is, 
\begin{equation}
C=\frac{(1+x)(1+x^{2})^{1/2}}{1+x+x^{2}},
\label{eqn:9}
\end{equation} 
which is nearly $0.936$ at origin and nearly unity for $|\alpha|^{2}\geq 3$ as shown in Fig.(\ref{fig:1}). We shall be working for coherent amplitudes appreciable enough so that the entanglement is nearly unity.

Alice mixes modes 0 and 1 in her possession to modes 3 and 4 using a symmetric lossless beam splitter, ${\cal B}_1$ fitted with two phase shifters at its second input and second output port which changes state $|\alpha\rangle$ to $|-i\alpha\rangle$. For coherent states, $|\alpha\rangle$ and $|\beta\rangle$ at the input ports $x$ and $y$, this combination transforms it to modes $u$ and $v$ as, $|{\alpha},{\beta}\rangle_{x,y}\rightarrow |\frac{{\alpha}+{\beta}}{\sqrt{2}},\frac{{\alpha}-{\beta}}{\sqrt{2}}\rangle_{u,v}$. Following \cite{prakash2007improving} we write,
 \begin{eqnarray}
 \label{eqn:10}
|\pm\sqrt{2}{\alpha}\rangle &=& \nonumber x|0\rangle+\frac{1-x^2}{\sqrt{2}}|{NZE,\sqrt{2}\alpha}\rangle \\  && \pm\sqrt{\frac{1-x^4}{2}}|{ODD,\sqrt{2}\alpha}\rangle,
 \end{eqnarray}
where, $|NZE,\sqrt{2}\alpha\rangle=[\sqrt{2}(1-x^{2})]^{-1}(|\sqrt{2}\alpha\rangle+|-\sqrt{2}\alpha\rangle-2x|0\rangle)$ is the normalized state containing non-zero even Fock states. We expand modes 3 and 4 of $|\psi\rangle_{3,4,2}$ in the orthogonal basis, \{$|0\rangle$, $|NZE,\sqrt{2}\alpha\rangle$, $|ODD,\sqrt{2}\alpha$\}. State of the system consisting of modes 3,4 and 2 can then be written as,
\begin{widetext}
\begin{eqnarray}
\label{eqn:11}
|\psi\rangle_{3,4,2} \nonumber 
&=& \frac{1}{\sqrt{2(1-x^{3}}}\left\{
x|0\rangle_{3}|0\rangle_{4}
\left( \epsilon_{+}+\epsilon_{-}\right)
\left( \left|\frac{\alpha}{\sqrt{2}}\right\rangle_{2}
+\left|-\frac{\alpha}{\sqrt{2}}\right\rangle_{2} 
\right)
+\frac{1-x^{2}}{\sqrt{2}}\left[ |NZE,\sqrt{2}\alpha\rangle_{3}|0\rangle_{4}\left(\epsilon_{+}\left|\frac{\alpha}{\sqrt{2}}\right\rangle_{2} 
\right.\right.\right.
\nonumber 
\\ &&
\left.\left.
 -\epsilon_{-}\left|-\frac{\alpha}{\sqrt{2}}\right\rangle_{2}\right)+
|0\rangle_{3}|NZE,\sqrt{2}\alpha\rangle_{4}\left(-\epsilon_{+}\left|-\frac{\alpha}{\sqrt{2}}\right\rangle_{2}+\epsilon_{-}\left|\frac{\alpha}{\sqrt{2}}\right\rangle_{2}\right)\right] 
+\sqrt{\frac{1-x^{4}}{2}}\left[|ODD,\sqrt{2}\alpha\rangle_{3}|0\rangle_{4}\right.
\nonumber 
\\ &&
\left. \left.
\left(\epsilon_{+}\left|\frac{\alpha}{\sqrt{2}}\right\rangle_{2}
+\epsilon_{-}\left|\frac{\alpha}{\sqrt{2}}\right\rangle_{2}\right)  
- |0\rangle_{3}|ODD,\sqrt{2}\alpha\rangle_{4}
\left(\epsilon_{+}\left|-\frac{\alpha}{\sqrt{2}}\right\rangle_{2}+\epsilon_{-}\left|-\frac{\alpha}{\sqrt{2}}\right\rangle_{2}\right)\right]\right\}.  
\end{eqnarray}
\end{widetext}
\begin{table*}[!hbt]
\begin{tabular}{|l|l|l|l|l|l|l|}
 \hline
Cases & Alice's PC Result & Probability & Bob's State in mode 2 (un-normalized) & U (Bob) & Mixing mode & BS-II output \\ \hline
 (i) & {0, 0} & $P_{i}$ & $|\alpha/\sqrt{2}\rangle-|-\alpha\/\sqrt{2}\rangle$ & - & - & - \\ \hline
 (ii) & {$NZE, 0$} & $P_{+}$ & $\epsilon_{+}|\alpha/\sqrt{2}\rangle-\epsilon_{-}|-\alpha\/\sqrt{2}\rangle$ & $I$ & $|ODD,\alpha/\sqrt{2}\rangle$ & $|I,0\rangle_{6,7}-|0,I\rangle_{6,7}$ \\ \hline
 (iii) & {$0, NZE$}  & $P_{+}$ & $-\epsilon_{+}|-\alpha/\sqrt{2}\rangle+\epsilon_{-}|+\alpha\/\sqrt{2}\rangle$ & $P(\pi)$ & $|ODD,\alpha/\sqrt{2}\rangle$ & $-|I,0\rangle_{6,7}+|0,I\rangle_{6,7}$ \\
 \hline
 (iv) & {$ODD, 0$} & $P_{-}$ &  $\epsilon_{+}|\alpha/\sqrt{2}\rangle+\epsilon_{-}|-\alpha\/\sqrt{2}\rangle$ & $I$ & $|EVEN,\alpha/\sqrt{2}\rangle$ & $|I,0\rangle_{6,7}+|0,I\rangle_{6,7}$ \\ \hline 
 (v) &  {$0, ODD$} & $P_{-}$ &  $-\epsilon_{+}|-\alpha/\sqrt{2}\rangle-\epsilon_{-}|\alpha\/\sqrt{2}\rangle$ & $P(\pi)$ & $|EVEN,\alpha/\sqrt{2}\rangle$ &  $-(|I,0\rangle_{6,7}+|0,I\rangle_{6,7})$ \\ 
 \hline 
\end{tabular}
\caption{Shows the result of photon counting done by Alice in modes 3 and 4, with corresponding probability of occurrence. The teleported state with Bob in mode 2, the conditional unitary operation applied by him, the mixing mode and the  entangled output state of ${\cal B}_2$ is also given. Here, $P_{+}$ and $P_{-}$ are given by Eqs (12) and (13), respectively, while $ I$ is the identity operation and $P(\pi)$ is phase shift by $\pi$.}
\label{table:1}
\end{table*}
\begin{figure*}[t] 
\includegraphics[width=\linewidth]{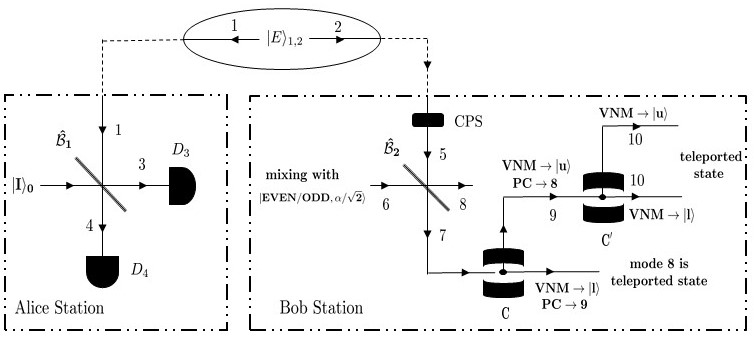}
\caption{Schematic of the proposed scheme for the teleportation of SCS using ECS. Alice mixes the information state, $|I\rangle_{0}$ in mode 0 with mode 1 of entangled state $|E\rangle_{1,2}$ using symmetric beam splitter (${\cal B}_1$) and performs photon counting measurement on the output modes 3 and 4 using detectors $D_1$ and $D_2$, respectively. She sends her measurement outcome to Bob using 2-bit classical channel, depending on which, Bob performs a conditional phase shifting (CPS) on mode 2. The CPS output mode 5 is further mixed with either $|ODD\alpha/\sqrt{2}\rangle|$, $|EVEN\alpha/\sqrt{2}\rangle|$, conditioned to classical result, with mode 6 using ${\cal B}_2$. The output modes of ${\cal B}_2$, 7 and 8, are the entangled mixture of vacuum and information.  Bob let mode 7 to interact with TLA initially in ground state $|l\rangle$, inside a cavity $C$. A VNM is performed on TLA in $C$, for $|l\rangle$, PC is done on 9 and mode 8 is teleported state. For $|u\rangle$, PC on mode 8 gives vacuum and mode 9 is again allowed to interact with second TLA initially in $|u\rangle$, inside cavity $C'$. VNM is again performed on TLA in $C'$, and the output mode 10 is taken as teleported state, irrespective of VNM result.} 
 \label{fig:4}
 \end{figure*} 
 
She then counts photons in modes 3 and 4, for which the detectors outcomes ($D_3,D_4$) can be one of the following, (0,0), (NZE,0), (0,NZE), (odd,0) or (0,odd) (cases (i) to (v)) and conveys this information to Bob.
For the case (i), when both detectors do not click, the collapsed state with Bob is $|ODD,\frac{\alpha}{\sqrt{2}}\rangle$ and it is not possible to transform this to information state using any means. However, the probability of occurrence is,
\begin{equation}
\label{eqn:12}
P_{i}=\frac{xP_{I0}}{1+x+x^2}
\end{equation}
which becomes almost zero for appreciable coherent amplitude $|\alpha|^{2}\geq 3 $ and for all $\theta$ (Fig.\ref{fig:2}). For information state $|\psi^I\rangle$ and teleported state $|\psi^T\rangle$ the teleportation fidelity is defined as,
\begin{equation}
\label{eqn:13}
F=|\langle \psi^I|\psi^T\rangle|^2.
\end{equation} 
For this case of photon counts (0,0) in the modes 3 and 4, the fidelity becomes,
\begin{equation}
\label{eqn:14}
F_i=\frac{x^{3/2-\sqrt{2}}(1-x^{\sqrt{2}})^2|\epsilon_{+}-\epsilon_{-}|^2}{2(1-x)}.
\end{equation} 

Denoting by $P_{+}$ and $P_{-}$ the probability of occurrence for cases (ii) (same for (iii)), and, case (iv) (same for (v)), respectively, we have,
\begin{equation}
P_{+}=\frac{(1+x)^2[1-P_{I0}]}{4(1+x+x^2)}
\label{eqn:15}
\end{equation}
\begin{equation}
\label{eqn:16}
P_{-}=\frac{(1+x^2)[1+P_{I0}]}{4(1+x+x^2)}.
\end{equation} 
$P_{+}$ and $P_{-}$ are plotted in Fig.(\ref{fig:3}) where both converges to a constant value of $1/4$ for appreciable coherent amplitudes, imitating the common SQT where all the four Bell states are equally probable.

It is now where our scheme differs from the previous ones (Fig.\ref{fig:4}). Alice communicates to Bob her measurement outcome using a classical channel. Bob on receiving the classical information, passes mode 2 through a conditional phase shifter (CPS), which transform it to mode 5. The CPS introduces a phase shift of $\pi$ for cases (ii) and (iv), and no phase shift for cases (i) and (iii). 
For cases (ii) and (iii), Bob mixes mode 5 with an odd coherent state $|ODD,\frac{\alpha}{\sqrt{2}}\rangle_{6}=[2(1-x)]^{-1/2}(|\frac{\alpha}{\sqrt{2}}\rangle_{6}-|-\alpha/\sqrt{2}\rangle_{6})$ in mode 6, using a symmetric beam splitter (${\cal B}_2$). The resulting state in modes 7 and 8 found to be, 
\begin{equation}
\label{eqn:18}
|\psi\rangle_{7,8} =\pm [2(1-P_{I0})]^{-1/2}(\pm|I,0\rangle_{7,8}- |0,I\rangle_{7,8}),
\end{equation}
respectively. On the other hand, for cases (iv) and (v), Bob mixes mode 5 with an even coherent state $|EVEN,\frac{\alpha}{\sqrt{2}}\rangle_{6}=2(1+x)]^{-1/2}(|\frac{\alpha}{\sqrt{2}}\rangle_{6}+(|-\frac{\alpha}{\sqrt{2}}\rangle_{6})$ leading to state in modes 7 and 8,
\begin{equation}
\label{eqn:19}
|\psi\rangle_{7,8} =\pm [2(1+P_{I0})]^{-1/2} (|I,0\rangle_{7,8} + |0,I\rangle_{7,8}),
\end{equation}
respectively. The state in eq.\eqref{eqn:18} and eq.\eqref{eqn:19} are an entangled mixture of vacuum and the information. Table \ref{table:1} shows the possible measurement outcomes of photon counting by Alice in modes 3 and 4, corresponding probability of occurrence, the state in mode 5, unitary transformation to be performed by Bob and the state with which Bob mixes mode 5 using ${\cal B}_2$ and the final state in modes 7 and 8.

From Table 1 it can be easily seen that apart from overall negative sign, that does not going to effect the foregoing analysis, cases (ii) and (iii), and cases (iv) and (v) gives similar entangled mixtures of information with vacuum in modes 7 and 8. In order to get perfect success, Bob needs to discriminate which of the mode contains the information state, and that too non-destructively.

Measurement in quantum mechanics brings an irreversible change in the state of the measured system due to the collapse of wavefunction.  Quantum non-demolition measurements is an interesting measurement scheme where the state of a system can be revealed without altering its state. Measuring certain properties of optical field, for example, photon number and optical quadratures, non-destructively, has been discussed in the past with many applications in quantum optics and quantum information science \cite{paris2001optimized,kok2002single}. However, the mode on which we are interested in making the measurement can be vacuum or the superposition of coherent state for which the trivial schemes of non-destructive measurement cannot be appropriate. Moreover, the information state $|I\rangle$ itself contains the vacuum, thus, this cannot be done with unit success. In the next section, we discuss a scheme of discriminating between vacuum state and the information, using the interaction of the mode of the light with a two-level atom (TLA). As we shall see, by entangling the mode with a TLA in a cavity and using atom as a probe, the desired task can be obtained with fidelity approaching unity with increasing $|\alpha|^{2}$.
\section{getting replica of information by using cavity-QED two exactly two-level atoms}
\label{sec:3}
\begin{figure*}[t]
\subfigure{
\label{fig:5a} 
\begin{minipage}[b]{0.45\linewidth}
\centering \includegraphics[width=3.32in]{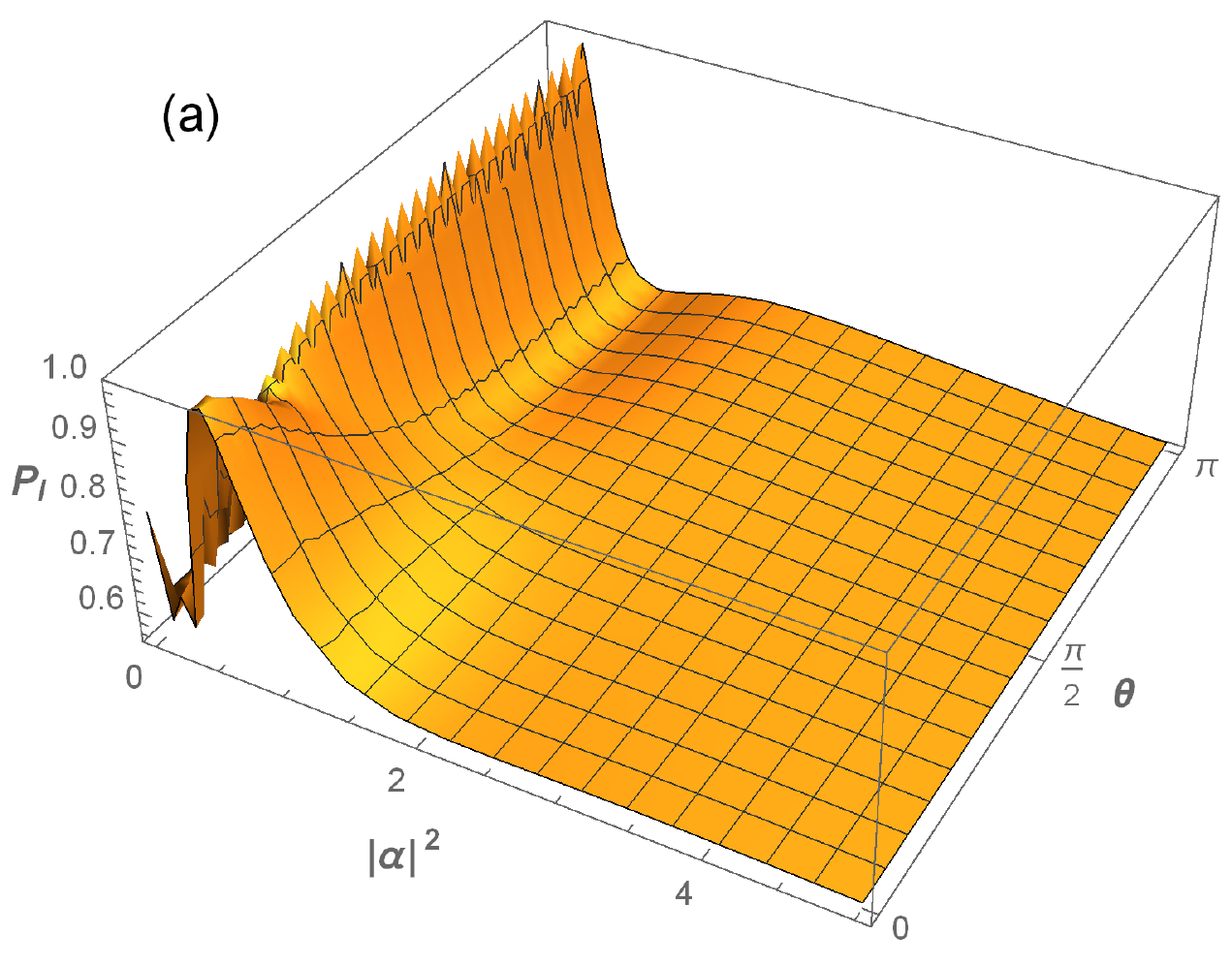}
\end{minipage}}%
\hspace{.12in}
\subfigure{
\label{fig:5b} 
\begin{minipage}[b]{0.45\linewidth}
\centering \includegraphics[width=3.32in]{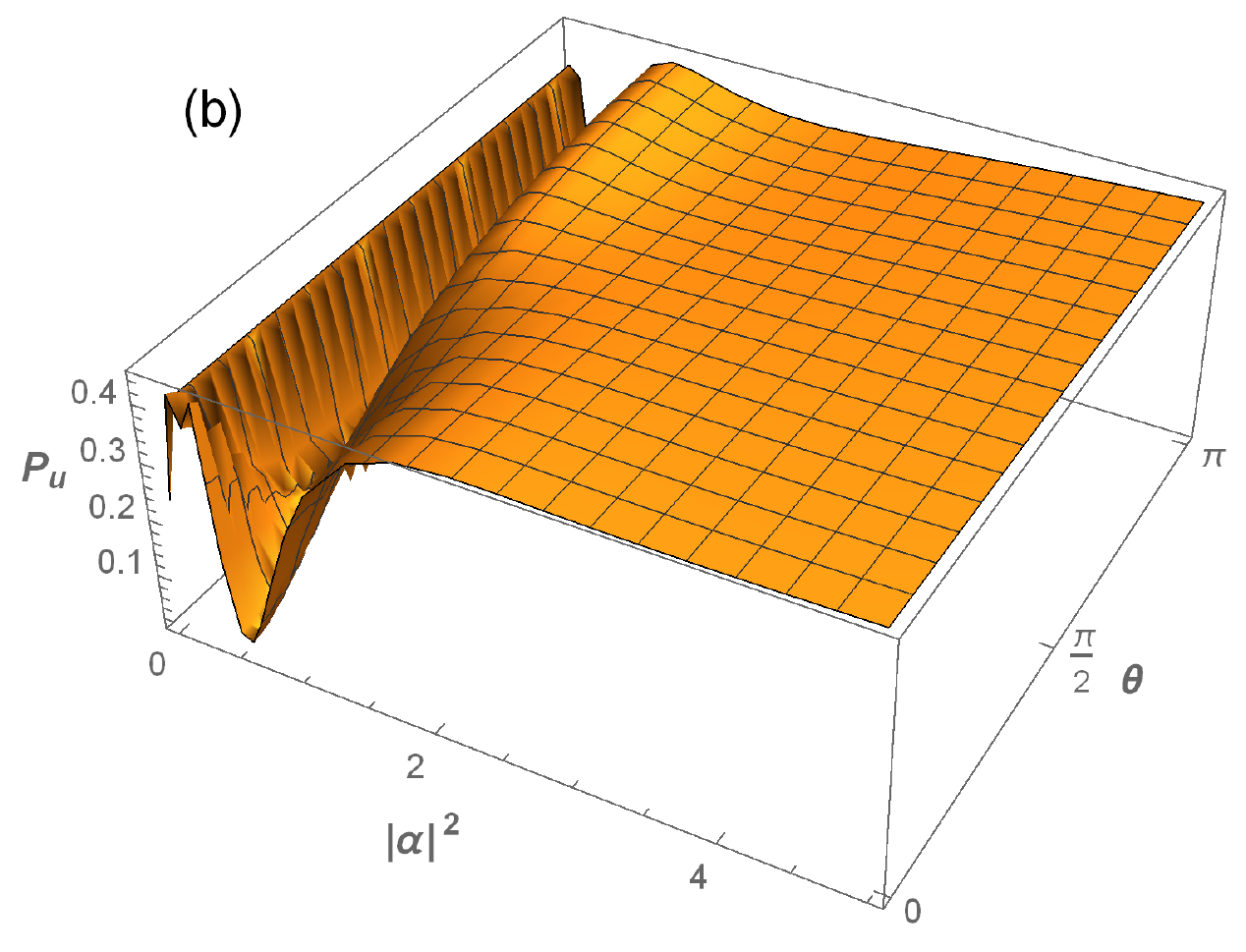}
\end{minipage}}
\subfigure{
\label{fig:5c} 
\begin{minipage}[b]{0.45\linewidth}
\centering \includegraphics[width=3.32in]{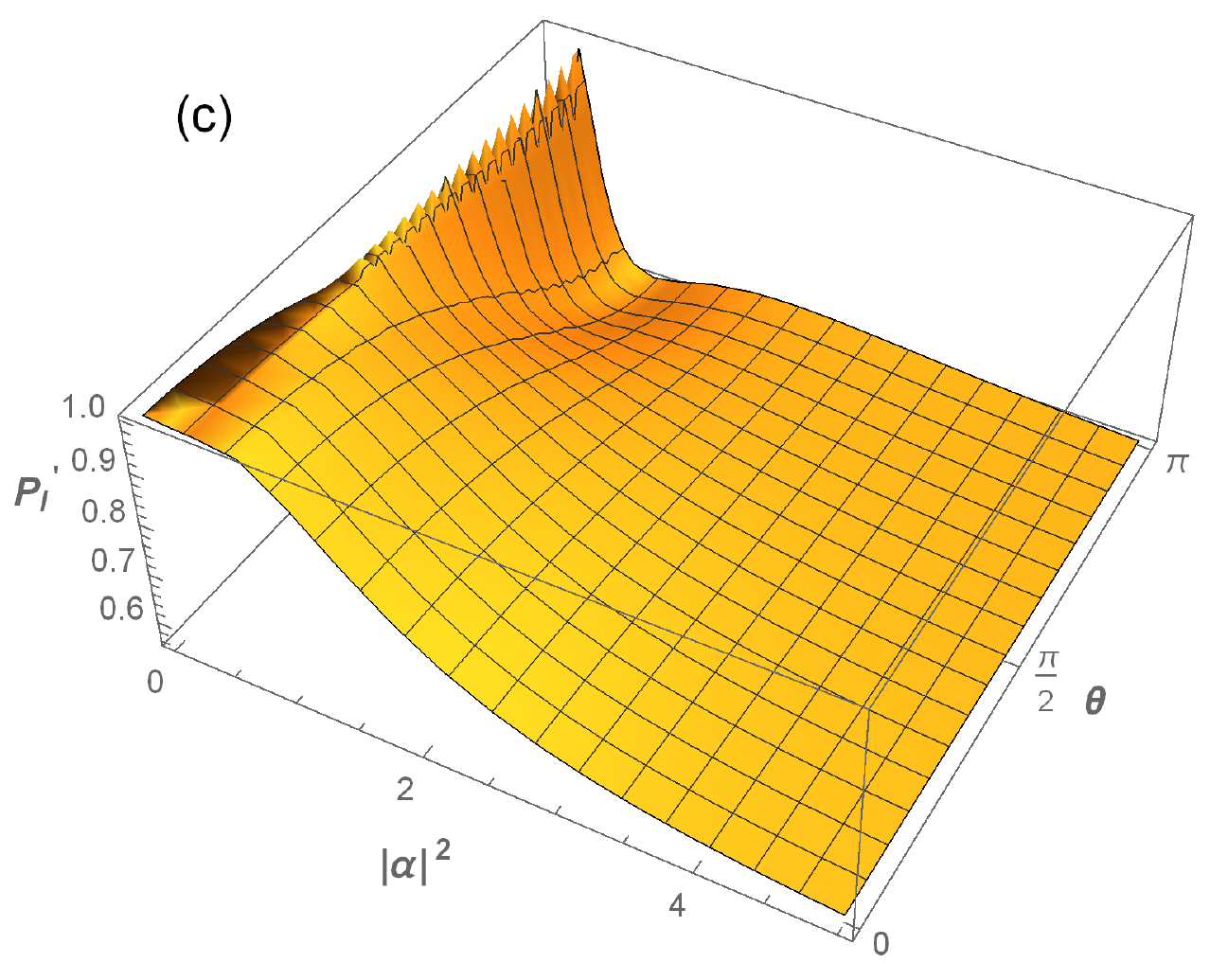}
\end{minipage}}%
\hspace{.12in}
\subfigure{
\label{fig:5d} 
\begin{minipage}[b]{0.45\linewidth}
\centering \includegraphics[width=3.32in]{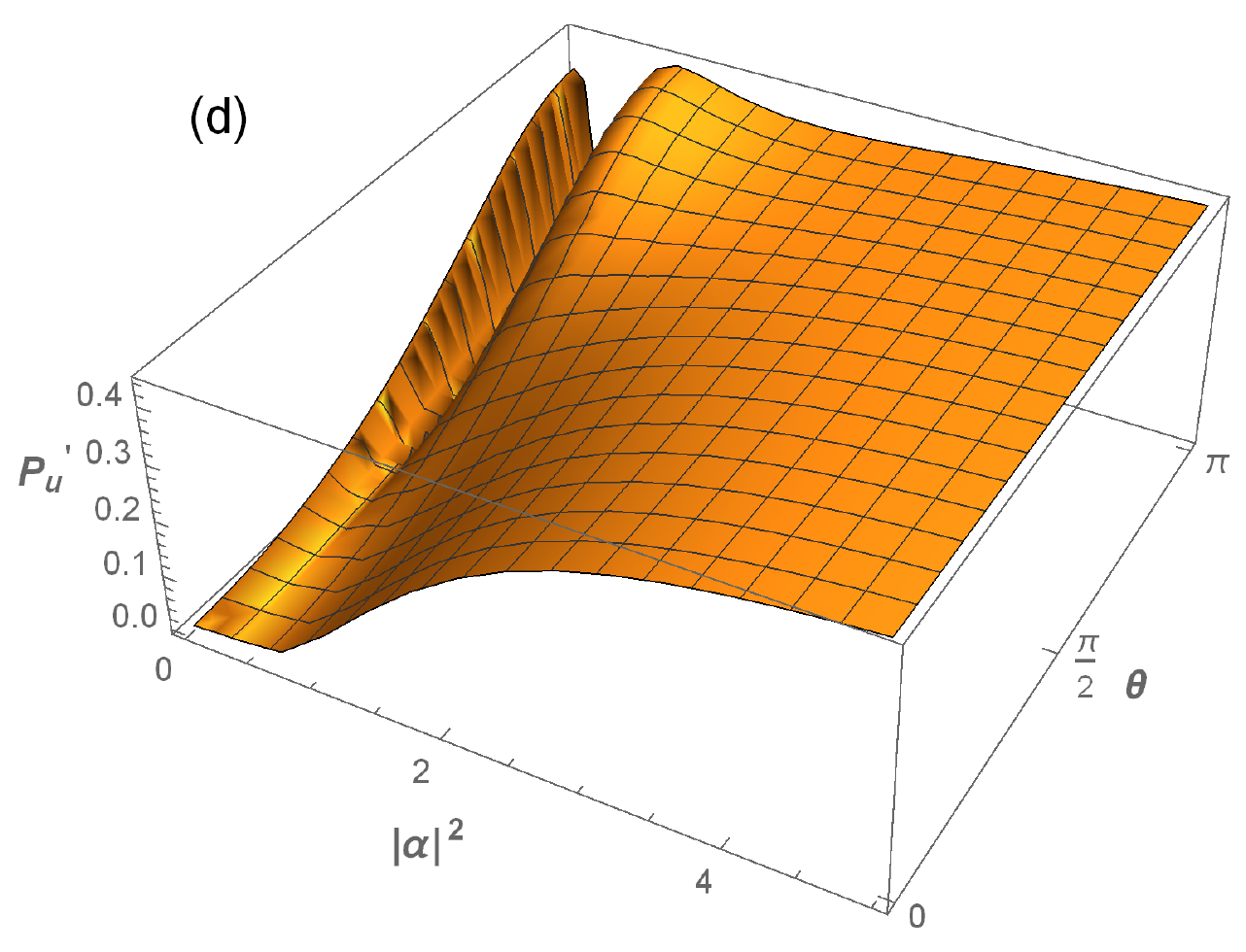}
\end{minipage}}
\caption{(a) and (b) shows the variation of probability for TLA in the cavity $C$ to be found in excited state and ground state for case (ii) (same for case (iii)) respectively, while (c) and (d) for case (iii) (same for case (iv)), with respect to information parameter $\theta$ and average photon number in the coherent state $|\alpha|^{2}$.}
\label{fig:5} 
\end{figure*}
\begin{figure*}[t]
\subfigure{
\label{fig:6a} 
\begin{minipage}[b]{0.45\linewidth}
\centering \includegraphics[width=3.32in]{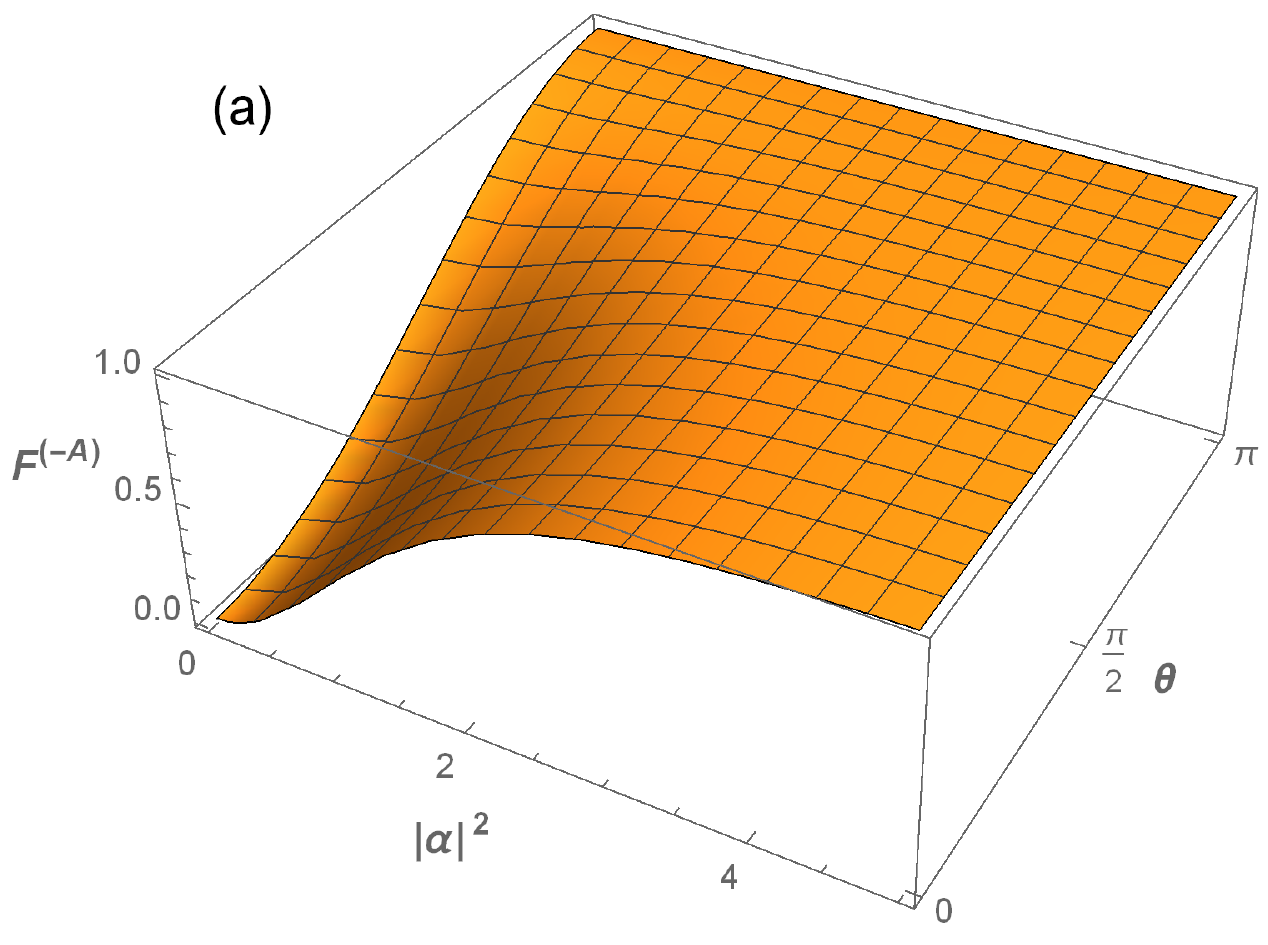}
\end{minipage}}%
\hspace{.12in}
\subfigure{
\label{fig:6b} 
\begin{minipage}[b]{0.45\linewidth}
\centering \includegraphics[width=3.32in]{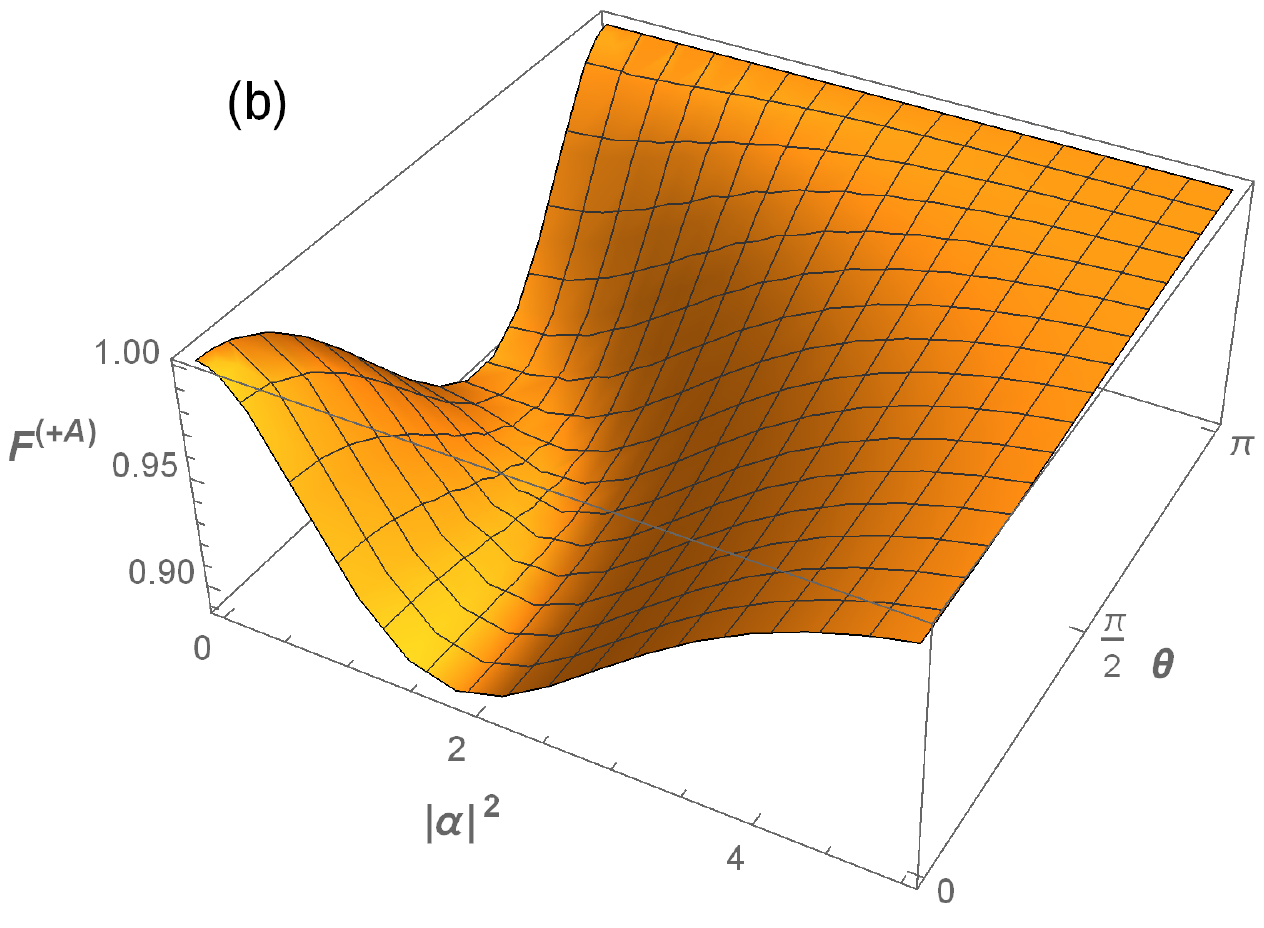}
\end{minipage}}
\caption{(a) and (b) shows the variation of fidelity for cases $A^-$ and $A^+$ with respect to $\theta$ and $|\alpha|^{2}$ respectively.}
\label{fig:6} 
\end{figure*}
Before we proceed to discuss our scheme to obtain a replica of information, let us first briefly discuss the dynamics of a single TLA (in lower state $|l\rangle$ or in excited state $|u\rangle$) interacting with a cavity mode in Fock state $|n\rangle$. The quantum mechanical interaction of cavity mode with TLA is best given by the Jaynes-Cumming model for which the interaction Hamiltonian can be written as,
\begin{equation} 
\label{eqn:20}
H_{int.}=\hbar g(a^{\dagger}\sigma_{-}+a\sigma_{+}),
\end{equation} 
where $a^{\dagger}(a)$ is the creation (annhilation) operator of the field mode, $g$ is the atom field coupling constant and $\sigma_{\pm}$ are the atomic raising and lowering operators \cite{gerry2005introductory}. Under such an interaction, the combined state of single cavity mode having 'n' photons and a TLA initially in its ground state $|l\rangle$ or in upper state $|u\rangle$, evolves as
\begin{equation} 
\label{eqn:21}
|n,l\rangle \longrightarrow \cos\phi_{n}|n,l\rangle-i\sin\phi_{n}|n-1,u\rangle,
\end{equation}
or
\begin{equation}
|n,u\rangle \longrightarrow \cos\phi_{n+1}|n,u\rangle-i\sin\phi_{n+1}|n+1,l\rangle
\end{equation}
where, $\phi_{n}=g\sqrt{n}t$. We use these to find the replica of information. 

Let Bob pass mode 8 to interact in a cavity $C$ with a resonant TLA in its ground state $|l\rangle$. If the changed state of radiation  is denoted by mode 9, the state of modes 7 and 9 and the TLA is
\begin{eqnarray}
\label{eqn:22}
|\psi^{\mp}\rangle_{7,9,C} &=& [2(1\mp P_{I0})]^{-1/2}[
(|I\rangle_{7}\mp p_{I0}|0\rangle_{7})|0\rangle_{9}|l\rangle_{C}
\nonumber \\  && \nonumber 
\mp \sum_{n=1}^{\infty}p_{In}|0\rangle_{7}(\cos\phi_{n}|n\rangle_{9}|l\rangle_{C} 
\\  && 
 -i\sin\phi_{n}|n-1\rangle_{9}|u\rangle_{C})].
\end{eqnarray}

The upper and lower signs in $\pm$ corresponds to cases (ii) or (iii), and for cases (iv) or (v), respectively here and later.
The interaction time is so chosen that, 
\begin{equation} \label{eqn:23}
t=t_{0},g|\alpha|t_{0}=\pi/2.
\end{equation}
A Von Neumann measurement (VNM) performed on TLA shall give us results $|l\rangle$ or $|u\rangle$ with conditional probabilities,
\begin{equation} \label{eqn:24}
P(\mp|l)=[2(1\mp P_{I0})]^{-1}\sum_{n=0}^{\infty}P_{In}\cos^{2}\phi_{n}+1\mp 2P_{I0}
\end{equation}
\begin{equation} \label{eqn:25}
P(\mp|u)=[2(1\mp P_{I0})]^{-1}\sum_{n=0}^{\infty}P_{In}\sin^{2}\phi_{n},
\end{equation}

Fig.\ref{fig:5} shows the variation of total probability of occurrence, $P_{l \: or \: u}=P_- P(-|l \: or \: u)$ and $P'_{l \: or \:u}=P_+ P(+|l\: or \:u)$, with respect to information parameter $\theta$ and mean photons $|\alpha|^{2}$ for VNM outcome to be $|l\rangle$ or $|u\rangle$, respectively. At small value 0f $|\alpha|$, the probability that TLA remain in the ground state is larger as there are only few photon to excite the atom. However, both becomes equal and converges asymptotically to a constant value $1/2$ for appreciably large coherent amplitudes.


Upon performing VNM, Bob makes a photon counting (PC) on one of the modes 7 or 9. This may result in situations for each of cases I to V:
\begin{description}
  \item[$\bullet$ $A^{\mp}$- TLA in $|l\rangle_{C}$, PC on 9 gives $n=0$] 
  \item[$\bullet$ $B_{n}^{\mp}$- TLA in $|l\rangle_{C}$, PC on 9 gives $n\neq 0$] 
  \item[$\bullet$ $C^{\mp}$- TLA in $|u\rangle_{C}$, n=0 in mode 7 is indicated]. 
\end{description}
We shall now discuss these situations separately.
\subsection{Situations $A^{\mp}$}
\label{sec:3.1}

The term corresponding to this case in $|\psi^{\pm}\rangle_{7,C,9}$ is given by,
 \begin{equation}
 \label{eqn:26} 
[2(1\mp P_{I0})]^{-1/2}(|I\rangle_{7}\mp p_{I0}|0\rangle_{7})|0\rangle_{9}|l\rangle_{C}.
\end{equation}
Bob accepts state in mode 7 as the teleported state, which is given by
\begin{equation}
\label{eqn:27}
|T^{(\mp, A)}\rangle= (1\mp P_{I0})^{-1/2}(|I\rangle_{7}\mp p_{I0}|0\rangle_{7}).
\end{equation} 
The conditional probabilities of occurrence and the fidelities for such cases are given by,
\begin{equation}
\label{eqn:28}
P(-|A)=\frac{1}{2}, P(+|A)=\frac{1+3P_{I0}}{2(1+P_{I0})}
\end{equation}
\begin{equation}
\label{eqn:29}
F^{(-,A)}=1-P_{I0}
$$$$
F^{(+,A)}=\frac{1+2P_{I0}+P_{I0}^{2}}{1+3P_{I0}}.
\end{equation}
$F^{(\mp, A)}$ becomes almost unity for $|\alpha|^{2}\geqslant3$ (Fig.\ref{fig:6}), 
thus, we can say that for these cases, an almost perfect teleportation is established between Alice and Bob.
\subsection{Situations $B_{n}^{\mp}$}
\label{sec:3.2}
Due to non-zero contribution of vacuum in the information, it may happen that TLA is not at all excited while interacting with the information state. It is this case for which we get non-zero PC result in mode 9 and TLA in its ground state. The corresponding term for these cases in $|\psi^{\pm}\rangle_{7,9,C}$ are,
\begin{eqnarray}\label{eqn:32}
[2(1\mp P_{I0})]^{-1/2}[
\mp p_{In}|0\rangle_{7}(\cos\phi_{n}|n\rangle_{9}|l\rangle_{C} 
\nonumber \\ 
 -i\sin\phi_{n}|n-1\rangle_{9}|u\rangle_{C})],
\end{eqnarray}
for $n\neq 0$. Since the teleported state with Bob is $|0\rangle_{7}$, the fidelity of teleportation  is,
\begin{equation}
\label{eqn:33}
F^{(\mp B_{n})}=P_{I0}.
\end{equation}
The conditional probability of occurrence for these cases becomes,
\begin{equation}
\label{eqn:34}
P(\mp |B_{n})= [2(1\mp P_{I0})]^{-1}\sum_{n=1}^{\infty}P_{In}\cos^{2}\phi_{n} .
\end{equation}
We note that $P(\mp |B_{n}) \leq 10^{-3}$ for $|\alpha|^2\geq 5$, therefore, such a case has negligibly small probability of occurrence for moderately large coherent amplitudes.
\subsection{Situations $C^{\pm}$}
\label{sec:3.3}
If VNM on TLA gives result $|u\rangle$, then it must be \textit{something} (at least one photon) that caused the excitation, ensuring that the mode 9 comes out of $|I\rangle$ while 7 is vacuum. The term corresponding to this case in $|\psi^{\mp}\rangle_{7,9,C}$ is given by,
\begin{equation}
\label{eqn:35}
-i[2(1\mp P_{I0})]^{-1/2}
\mp \sum_{n=1}^{\infty}p_{In}|0\rangle_{7}(
 \sin\phi_{n}|n-1\rangle_{9}|u\rangle_{C})].
\end{equation}
However, due to absorption of photon from mode 8 in exciting the atom, the state (information) is changed. To reverse this effect (atleast approximately), we pass mode 9 to another cavity ($C^{'}$) where TLA is initially in $|u\rangle$, and let it interact for another time $t_{0}$. Here, we are taking advantage of the phenomenon of collapses and revivals of atomic population when a TLA is allowed to interact with a cavity mode which is in superposition of Fock states (for our case, the information state).
The interaction of mode 9 with second cavity $C'$ transforms as
\begin{equation}
\label{eqn:36}
|n-1\rangle_{9}|u\rangle_{C'}\rightarrow\cos\phi_{n}|n-1\rangle_{10}|u\rangle_{C'}-i\sin\phi_{n}|n\rangle_{10}|l\rangle_{C'}.
\end{equation}
A VNM is again performed on $C'$, resulting in following two cases. We note that, except the  expression for conditional probabilities, fidelity as well as teleported state shall remain same for cases (ii) to (v).

\subsubsection{ When VNM in C' gives $|l\rangle$}
\label{sec:3.3.1}
Using Eq. (36) and (37), the term corresponding to these cases are,
\begin{eqnarray}
\label{eqn:37}
 -[2(1\mp P_{I0})]^{-1/2}  
\sum_{n=1}^{\infty}p_{In}
 \sin^{2}\phi_{n}|n\rangle_{10}|l\rangle_{C'},
\end{eqnarray}
with respective conditional probability of occurrence given by,
\begin{equation}
\label{eqn:38}
P(\mp |C_{u})= [2(1\mp P_{I0})]^{-1}\sum_{n=1}^{\infty}P_{In}\sin^{4}\phi_{n}.
\end{equation}
The teleported state in mode 10 is,
\begin{equation}
\label{eqn:39}
|T^{(C_{l}})\rangle=\frac{\sum_{n=1}^{\infty}p_{In}\sin^{2}\phi_{n}|n\rangle}{\sqrt{\sum_{n=1}^{\infty}P_{In}\sin^{4}\phi_{n}}},
\end{equation}
therefore, the fidelity of teleportation becomes,
\begin{equation}
\label{eqn:40}
F^{(C_{l})}=\frac{[\sum_{n=1}^{\infty}P_{In}\sin^{2}\phi_{n}]^{2}}{\sum_{n=1}^{\infty}P_{In}\sin^{4}\phi_{n}}.
\end{equation}
We have numerically plotted $F^{(C_{l})}$ with repect to $|\alpha|^2$, and information parameters $\theta$ and $\phi$. The fidelity becomes almost unity for $|\alpha|^2\geq 3$.
\subsubsection{When VNM in C' gives $|u\rangle$}
\label{sec:3.3.2}
\begin{figure}
  \includegraphics[width=\linewidth]{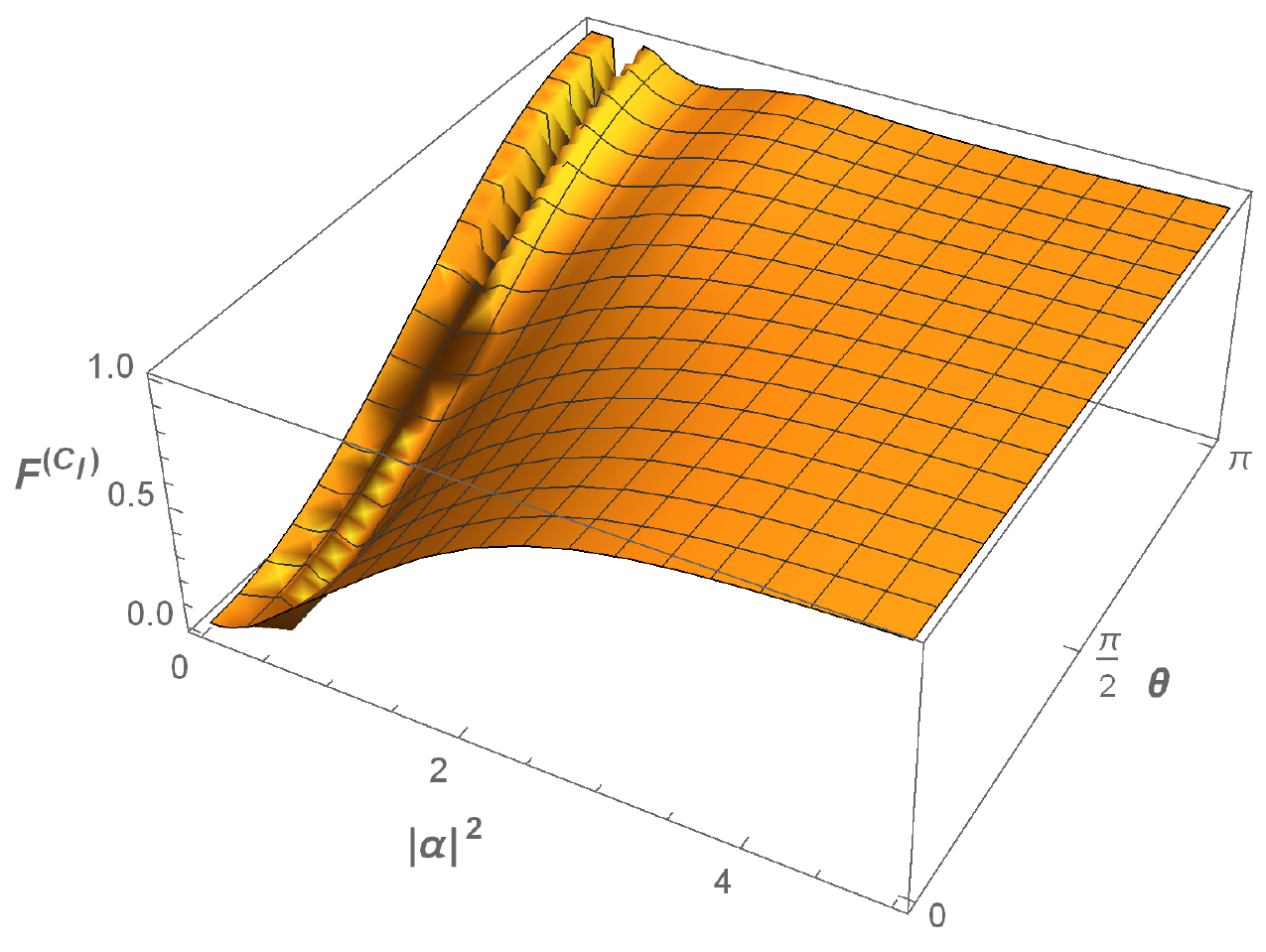}
  \caption{Variation of fidelity $F^{(C_l)}$ with respect to $\theta$ and $|\alpha|^{2}$. The fidelity becomes nearly unity as mean photons the coherent state increases and for all $\theta$}
   \label{fig:7}
\end{figure}
There is still a non-zero probability that the TLA remains excited despite of the interaction with mode 9, containing the vacuum state. The term corresponding to these cases are,
\begin{eqnarray}\label{eqn:41}
 -[2(1\mp P_{I0})]^{-1/2}
\sum_{n=1}^{\infty}p_{In}
 \sin\phi_{n}\cos\phi_{n}|n-1\rangle_{10}|u\rangle_{C'},
\end{eqnarray}
with respective conditional probabilities,
\begin{equation}
\label{eqn:42}
P(\mp |C_{u})= [2(1\mp P_{I0})]^{-1}\sum_{n=1}^{\infty}P_{In}\sin^{2}\phi_{n}\cos^{2}\phi_{n}.
\end{equation}
The teleported state with Bob in mode 10 is,
\begin{equation}
\label{eqn:43}
|T^{(C_{u})}\rangle=\frac{\sum_{n=1}^{\infty}p_{In}\sin\phi_{n}\cos\phi_{n}|n-1\rangle}{\sqrt{\sum_{n=1}^{\infty}P_{In}\sin^2\phi_{n}\cos^2\phi_{n}\phi_{n}}}.
\end{equation}
This gives the fidelity,
\begin{equation}
\label{eqn:44}
F^{(C_{u}})=\frac{|\sum^{\infty}_{n=1}p^*_{In-1}p_{In}\sin\phi_{n}\cos\phi_{n}|^2}{\sum_{n=1}^{\infty}P_{In}\sin^2\phi_{n}\cos^2\phi_{n}\phi_{n}}.
\end{equation}

\section{Overall quality of teleportation}
\label{sec:4} 
\begin{figure}[t]
  \includegraphics[width=\linewidth]{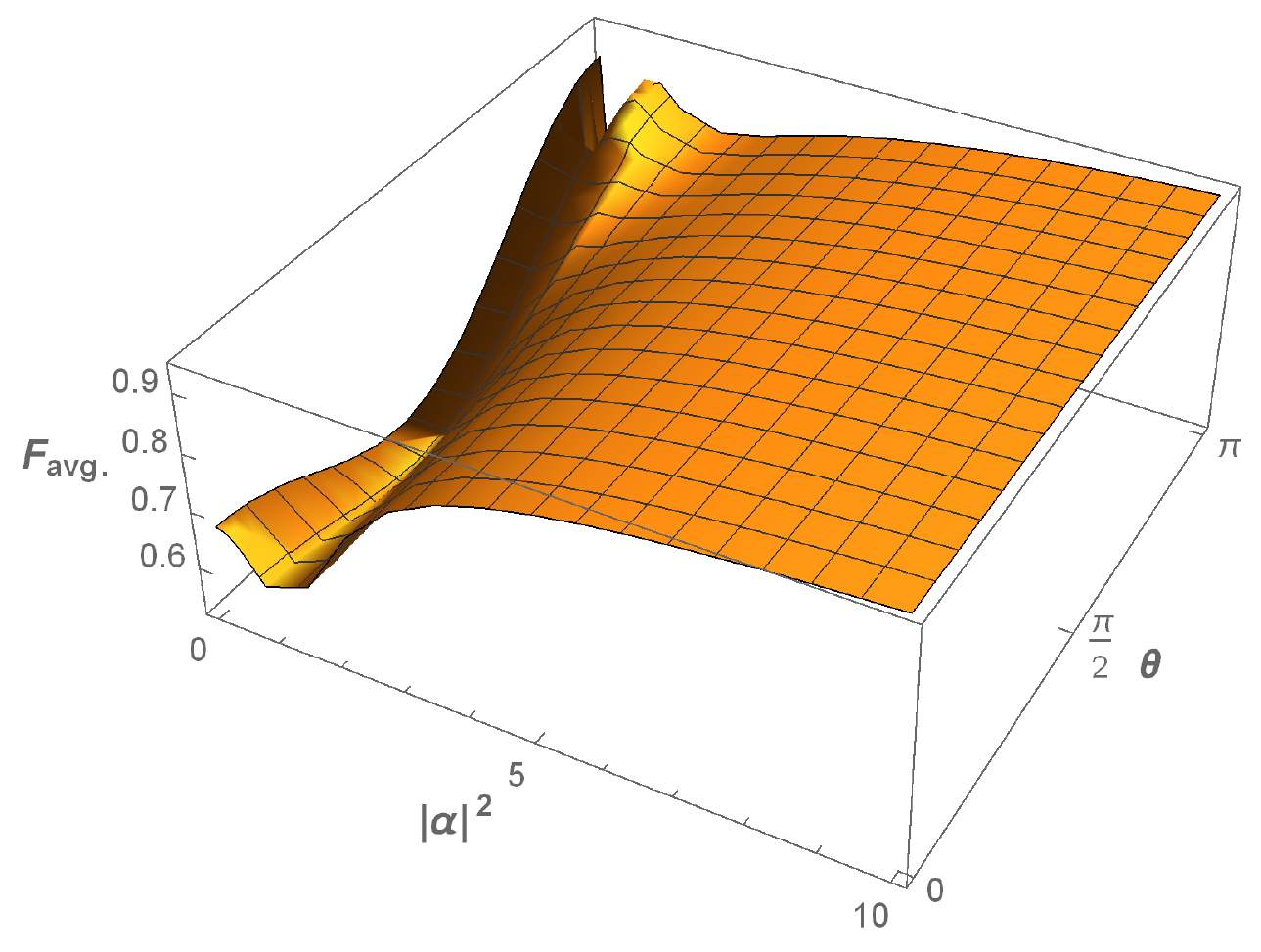}
  \caption{Variation of average fidelity of teleportation for the proposed scheme with respect to $\theta$ and $|\alpha|^{2}$. $F_{avg.}$ becomes nearly unity as mean photons the coherent state increases and for all $\theta$}
   \label{fig:8}
\end{figure}
To estimate overall quality of teleportation, we calculate the average fidelity of our teleportation scheme. Following Prakash et al. \cite{prakash2013quantum}, we define the average fidelity of teleportation as the sum of products of probability of occurrence with corresponding fidelity for each possible outcomes. Therefore, we can write average fidelity for our teleporation scheme as,
\begin{eqnarray} 
\label{eqn:45}
F_{avg.} &=& \nonumber 
P_{i}F_{i}+2P_{-}[P(-|A)F^{(-A)}+P(-|B_{n})F^{(-B_{n})}\\ &&+P(-|C_{l})F^{(C_{l})}+P(-|C_{u})F^{(C_{u})}]
\nonumber \\ &&
+2P_{+}[P(+|A)F^{(+A)}+P(+|B_{n})F^{(+B_{n})}
\nonumber \\ &&
+P(+|C_{l})F^{(C_{l})}+P(+|C_{u})F^{(C_{u})}].
\end{eqnarray}
By substituting the values of various terms, and on simplifying, we get,
\begin{eqnarray} 
\label{eqn:46}
F_{avg.} &=& \nonumber 
\frac{1}{2(1-x^3)]}\left[x^{2-3/2-\sqrt{2}}(1-x^{\sqrt{2}})^2|\epsilon^{2}_{+}-\epsilon^{2}_{-}|^2\right.
\\ && \nonumber
\left. -x(1-x)P_{I0} \right] +\frac{1}{2}\left[1+P^{2}_{I0}+P_{I0}S_1 \right.
\\ &&
\left. + |S_2|^2+|S_3|^2\right] .
\end{eqnarray}
where, 
\begin{equation}
\begin{gathered}
S_1=\sum_{n=1}^{\infty}P_{In}\cos^{2}\phi_{n},
\\
S_2=\sum_{n=1}^{\infty}P_{In}\sin^{2}\phi_{n},
\\
S_3=\sum^{\infty}_{n=1}p^*_{In-1}p_{In}\sin\phi_{n}\cos\phi_{n}.
\end{gathered}
\end{equation}
The average fidelity is a function of mean coherent amplitude $|\alpha|^{2}$ as well as information parameters, $\theta$ and $\phi$ which is plotted numerically in Fig. ({\ref{fig:8}}). We need to find analytically the dependence of $F_{avg.}$ over these independent parameters in order to analyse such a variation more closely. 

The summation terms, $S_1, S_2$ and $S_3$ follows from the Appendix. Using Eqs. \eqref{a:12}, \eqref{a:13} and \eqref{a:18} we can write the average fidelity,
\begin{center}
 \begin{widetext}
\begin{eqnarray} 
\label{eqn:47}
F_{avg.} &=& \nonumber 
1+\frac{1}{2(1-x^3)}\left[x^{2-3/2-\sqrt{2}}(1-x^{\sqrt{2}})^2|\epsilon^{2}_{+}-\epsilon^{2}_{-}|^2-x(1-x)P_{I0}\right]
+\frac{1}{2}\left\{ P^{2}_{I0}+\frac{P_{I0}\pi^2}{32}\left[\frac{2}{|\alpha|^2}-\frac{1}{|\alpha|^4}\right.\right.
\\ && \nonumber
\left. + \left( 8-\frac{5}{|\alpha|^2}
+\frac{1}{|\alpha|^4}\right)X\right]-P_{0I}-\frac{\pi^2}{8|\alpha|^2}+\frac{\pi^2(\pi^2+8)}{128|\alpha|^4}-\frac{\pi^2}{2}\left[1-\frac{(10+\pi^2)}{16|\alpha|^2}-\frac{(16+9\pi^2)}{16|\alpha|^4}\right] X
\\ &&\nonumber 
\left. + \pi^2 \left[ 2-\frac{1}{4|\alpha|^2}+\frac{41}{32|\alpha|^4}\right] X^2 \right\}+\frac{x^2\pi^2|k_1|^2}{256|\alpha|^4}-\frac{x^2\pi^2|k_2|^2}{144|\alpha|^2}\left[(\pi^2+3)^2-\frac{2(\pi^4+10\pi^2+21)}{|\alpha|^2} \right.
\\ &&
\left. +\frac{(\pi^4+9\pi^2+18)}{4|\alpha|^4}\right]+\frac{x^2\pi^2(k^*_1 k_2+k_1k_2^*)}{192|\alpha|^2}\left[(\pi^2+3)-\frac{(\pi^4+15\pi^2+60)}{6|\alpha|^2}\right],
\end{eqnarray}
\end{widetext}
 \end{center}
where,
\begin{equation}
\begin{gathered}
X=\frac{2x^2(|\epsilon_+ + \epsilon_-|^2-1)}{1-x^2},
\\
k_1=|\epsilon_+|^2-|\epsilon_-|^2
\\
k_2=x^2(\epsilon^*_+\epsilon_--\epsilon_+\epsilon^*_-).
\end{gathered}
\end{equation} 
At small coherent amplitude $|\alpha|^{2}\rightarrow 0$, considerable variation of $F_{avg.}$ with respect to $\theta$ is observed, maximizing for $\theta = \pi$. The reason for this is that the terms having $\theta$ dependence (terms containg $\epsilon_{\pm}$) are multiplied by either $x$ or its square, which contributes considerably in such limits. However, $x$ decreases very rapidly to become zero for $|\alpha|^2\gtrsim 3$ beyond which the variation ceases to exist. Also, the lower value of $F_{avg.}$ may be attributed to lower values of concurrence of $|E'\rangle_{1,2}$ and much higher probability for the case (i) to occur, compared to other cases. 

For coherent amplitude large enough, say $|\alpha|^2\geq5$, we can afford to discard terms proportional to $x$ and $X$. We can then write the average fidelity in this limit as,
\begin{equation}
\label{eqn:48}
F_{avg.}\simeq 1-\frac{\pi^2}{16|\alpha|^2}+\frac{\pi^2(\pi^2+8)}{256|\alpha|^4},
\end{equation}
giving $F_{avg.}=\{0.947, 0.971, 0.980\}$ at $|\alpha|^2=\{10, 20, 30\}$ respectively, and increasing monotonically to become unity. Therefore, we can say that our scheme gives an almost perfect quantum teleportation of SCS and it becomes more and more effective as for large coherent amplitudes.

\section{Conclusion}
\label{sec:5}
We presented a scheme for quantum teleportation of SCS using linear optics and photon detectors. The receiver only requires to perform either no operation or a $\pi$ phase shift, as the only form of unitary operation which is conditioned to the PC result of the sender. Furthermore, mixing of the unitary transformed state with either even or odd coherent state using a symmetric lossless beam splitter gives directly the entangled mixture of information with vacuum as the output. This entangled mixture ensures that only one of the output mode contains the information and other is a vacuum. Our scheme relies on using two exactly resonant TLA in cavities to obtain the replica of information. It is, however, demanding to find out more reliable method for exact extraction of $|I\rangle$, which is still open. 

Our scheme is experimentally feasible using current technological capabilities and it can be found useful in various quantum information processing tasks that involves coherent states and its entangled counterparts. The robustness of ECS against noise makes it an important choice for establishing long-distance quantum teleportation as well as for other quantum information processing tasks for which our scheme can be found useful.
\appendix*
\section{approximating $S_1$, $S_2$ and $S_3$}
\label{sec:appendix}
We need to find the summation of terms, $\sum^{\infty}_{n=1}P_{In}\cos^{2}\phi_{n}$, $\sum^{\infty}_{n=1}P_{In}\sin^{2}\phi_{n}$, $\sum^{\infty}_{n=1}p*_{In-1}p_{In}\sin\phi_{n}\cos\phi_{n}$, which we abbreviated by $S_1$, $S_2$ and $S_3$. We expand $\phi_{n}$ about $n=|\alpha|^2$ as,
\begin{equation}
\label{a:1}
\phi_n=\frac{\pi}{2}\sqrt{1+y}\simeq \frac{\pi}{2}\left( 1+\frac{y}{2}-\frac{y^2}{8}\right)
\end{equation}
where, $y=\frac{\delta n}{|\alpha|^2}$ and $\delta n=n-|\alpha|^2$. As $y<<1$, we shall only keep terms upto cubic order in $|\alpha|$. We shall further expand the trigonometric function to obtain,
\begin{equation}
\begin{gathered}
\label{a:2}
\sin\phi_{n}=1-\frac{\pi^{2}y^2}{32}+\frac{\pi^{2}y^3}{64}
\\
\cos\phi_{n}=-\frac{\pi y}{4}+\frac{\pi y^2}{16}+\frac{\pi^{3}y^3}{384}
\end{gathered}
\end{equation}
which can be used further to obtain,
\begin{equation}
\label{a:4}
\sin^2\phi_{n}=1-\frac{\pi^{2}y^2}{16}+\frac{\pi^{3}y^3}{32},
\end{equation}
\begin{equation}
\label{a:5}
\cos^2\phi_{n}=\frac{\pi^2 y^2}{16}-\frac{\pi^{2}y^3}{32},
\end{equation}
and
\begin{equation}
\label{a:6}
\sin\phi_{n}\cos\phi_{n}=-\frac{\pi y}{4}+\frac{\pi y^2}{16}+\frac{\pi^{3}y^3}{96}.
\end{equation}
We note that $P_{In}$ is sharlply peaked at $n=|\alpha|^2$, while the trigonometric fuction are flat and hence they can be replaced by their approximate values near $n=|\alpha|^2$. Therefore, finding the sum is reduced into calculating terms, $\sum_{n=0}^{\infty}y^m P_{nI}$, for $m=1,2$ and $3$. We use the fact that,
\begin{equation}
\label{a:7}
\sum_{n=0}^{\infty}f(n)P_{nI}=\langle I|f(N)|I\rangle,
\end{equation}
where $N=a^{\dagger}a$ is the number operator , together with,
\begin{equation}
\begin{gathered}
\label{a:8}
\langle I|a^{\dagger 2m}a^{2m}|I\rangle=|\alpha|^{4m}\\
\langle I|a^{\dagger 2m+1}a^{2m+1}|I\rangle=|\alpha|^{4m}(1-X),
\end{gathered}
\end{equation}
where, $X=\frac{2x^2(|\epsilon_{+}+\epsilon_{-}|^2-1)}{1-x^2}$, to obtain, 
\begin{equation}
\begin{gathered}
\label{a:9}
\sum_{n=0}^{\infty}yP_{nI}=X,
\\
\sum_{n=0}^{\infty}y^2 P_{nI}=\frac{1}{|\alpha|^2}+\left(2-\dfrac{1}{|\alpha|^2}\right) X,
\\
\sum_{n=0}^{\infty}y^3P_{nI}=\dfrac{1}{|\alpha|^4}+\left(-4+\dfrac{3}{|\alpha|^2}-+\dfrac{1}{|\alpha|^4}\right) X.
\end{gathered}
\end{equation}
We thus have,
\begin{eqnarray}
\label{a:12}
S_1 &=& \nonumber 
\frac{\pi^2}{32}\left[ \frac{2}{|\alpha|^2}-\frac{1}{|\alpha|^4}+ \left(8-\dfrac{5}{|\alpha|^2}+\dfrac{1}{|\alpha|^4}\right) X\right] \\ &&-P_{0I},
\end{eqnarray}
and,
\begin{eqnarray}
\label{a:13}
S_2 &=& 1-(T_{1}+P_{0I}).
\end{eqnarray}
A Little more effort is required to find $S_3$. We note that,
\begin{eqnarray}
\label{a:14}
S_3
 &=& \nonumber \frac{x}{\alpha*}\sum^{\infty}_{n=1}
Q(n)\sqrt{n}[|\epsilon_{+}|^2-|\epsilon_{-}|^2
\\  &&
+(-1)^n(\epsilon^*_+\epsilon_--\epsilon_+\epsilon^*_-)]\cos\phi_n\sin\phi_n
 \end{eqnarray}
where, $Q(n)=x|\alpha|^2/n!$. We can write, $\sqrt{n}=|\alpha|(1+y/2-y^2/8)$ upto second order in $y$. We use the fact that, 
\begin{equation}
\begin{gathered}
\label{a:15}
\langle\alpha|f(N)|\alpha\rangle=\sum_{n=0}^{\infty}Q(n)f(n)
\\
\langle\alpha|f(N)|-\alpha)=\sum_{n=0}^{\infty}(-1)^n Q(n)f(n)
\end{gathered}
\end{equation} 
which on putting in Eq. \eqref{a:14} and using 
\begin{equation}
\begin{gathered}
\label{a:16}
\langle -\alpha|a^{\dagger 2m}a^{2m}|\alpha\rangle=x^2|\alpha|^{4m}
\\
\langle -\alpha|a^{\dagger 2m+1}a^{2m+1}|\alpha\rangle=x^2|\alpha|^{4m+2}
\end{gathered}
\end{equation}
we obtain,
\begin{eqnarray}
\label{a:18}
S_3 &=& \nonumber
 \frac{\alpha}{|\alpha|}\left\{\frac{\pi(|\epsilon_{+}|^2-|\epsilon_{-}|^2)}{16|\alpha|^2}\left(-1+\frac{\pi^2+6}{6|\alpha|^2}\right) \right.
\\  && \nonumber
-\frac{\pi x^2(\epsilon_+*\epsilon_--\epsilon_+\epsilon_-*)}{12}\left[\pi^2+3-\frac{3(\pi^2+7)}{4|\alpha|^2} \right.
\\  &&
\left. \left. +\frac{\pi^2+6}{8|\alpha|^4}\right] \right\} 
 \end{eqnarray}
\begin{acknowledgements}
One of the authors R.K.P. is thankful to UGC for providing financial assistance under CSIR- UGC JRF fellowship. Discussions with Dr. Devendra Kumar Mishra and Ms. Shamiya Javed are gratefully acknowledged.
\end{acknowledgements}
\bibliographystyle{unsrt}
 
\end{document}